# Architecture de médiation « Tout-XML »

## Conception et évaluation

**Tuyet-Tram Dang-Ngoc — Georges Gardarin**

*Laboratoire PRiSM – Université de Versailles Saint Quentin*
*45, avenue des Etats-Unis*
*78035 Versailles Cedex*
*{Tuyet-Tram.Dang-Ngoc, Georges.Gardarin}@prism.uvsq.fr*

*RÉSUMÉ. XML s'est imposé comme le méta-langage permettant de représenter et d'échanger des données non seulement sur le web mais aussi de façon générale en entreprise. XQuery s'impose comme le langage de requête standard pour XML. En conséquence, des outils sont nécessaires pour interroger des sources de données hétérogènes avec XQuery, et ainsi intégrer des données hétérogènes en temps réel sur demande. Cet article présente le médiateur XMedia, un outil permettant d'intégrer et d'interroger des informations hétérogènes distribuées sous la forme de vues XML unifiées. Il décrit l'architecture du médiateur et se concentre sur la technique d'analyse de requêtes distribuées qui a été implémentée dans ce composant. L'évaluation de requête est basée sur une algèbre XML étendant simplement les opérateurs classiques de l'algèbre relationnelle à des traitements de tuples d'éléments arborescents. De plus, nous présentons un ensemble d'évaluation de performances sur un banc d'essai de type relationnel distribué, ce qui conduit à discuter d'éventuelles futures optimisations.*

*ABSTRACT. XML has emerged as the leading language for representing and exchanging data not only on the Web, but also in general in the enterprise. XQuery is emerging as the standard query language for XML. Thus, tools are required to mediate between XML queries and heterogeneous data sources to integrate data in XML. This paper presents the XMedia mediator, a unique tool for integrating and querying disparate heterogeneous information as unified XML views. It describes the mediator architecture and focuses on the unique distributed query processing technology implemented in this component. Query evaluation is based on an original XML algebra simply extending classical operators to process tuples of tree elements. Further, we present a set of performance evaluation on a relational benchmark, which leads to discuss possible performance enhancements.*

*MOTS-CLÉS : XML, architecture de médiation, algèbre XML, XQuery.*

# 1. Introduction

Ces dernières années ont vu apparaître beaucoup de projets de recherche basés sur l'intégration d'informations hétérogènes. Les systèmes typiques d'intégration d'information ont adopté une architecture médiateur-adaptateurs (Wiederhold, 1993). Dans cette architecture, les médiateurs fournissent une interface utilisateur uniforme pour interroger des vues intégrées de sources d'informations hétérogènes. Les adaptateurs fournissent des vues locales des sources de données dans un modèle global de données. Les vues locales peuvent être interrogées d'une manière limitée selon les possibilités des adaptateurs. Bien que l'approche *local-as-view* (LAV) ait été considérée dans quelques systèmes (Levy *et al.*, 1996) (Manolescu *et al.*, 2001), la plupart des systèmes suivent l'approche *global-as-view* (GAV), dans lesquelles les vues intégrées sont conçues en termes de vues locales des sources. Les projets de recherche et les prototypes les plus connus basés sur cette architecture incluent GARLIC (Haas *et al.*, 1997), TSIMMIS (Chawathe *et al.*, 1994), IRO-DB (Fankhauser *et al.*, 1998) et YaT (Cluet *et al.*, 1998).

Alors que dans les années 90 la plupart des études étaient basées sur le modèle objet comme modèle d'intégration de données, la recherche s'est concentré sur XML en tant que modèle global au début de ce nouveau siècle. Les avantages de XML comme modèle d'échange, (c.-à-d., riche, clair, extensible et sûr), le placent comme le meilleur candidat pour supporter les modèles de données intégrées. En outre, employer des vues XML pour des sources de données locales permet d'occulter les spécificités locales de chaque système. De plus, la richesse du modèle de schéma XML (XML-Schema) simplifie le travail des adaptateurs. Enfin, l'apparition de XQuery comme puissant langage d'interrogation universel pour XML permet d'interroger des vues globales et locales XML d'une manière uniforme basée sur une interface standard. Ainsi, tous ces avantages expliquent que plusieurs projets de recherche ont émergé pour interroger des sources des données hétérogènes d'une manière uniforme en se basant sur le modèle d'échange XML, par exemple (Christophides *et al.*, 2000), (Manolescu *et al.*, 2001) et (Shanmugasundaram *et al.*, 2001).

XMedia est l'un des premiers systèmes basés sur XML permettant d'intégrer des sources de données hétérogènes. Une première version a été développée à l'Université de Versailles (laboratoire PRiSM) puis industrialisée par la start-up e-XMLMedia (voir *http://www.e-xmlmedia.fr/*). Le système est actuellement disponible en logiciel libre (open source). Ce médiateur et les adaptateurs associés fournissent les fonctionnalités nécessaires pour interroger des sources de données hétérogènes de manière uniforme avec XML et XQuery. C'est un composant complexe composé de plusieurs paquetages responsables: 1- des fonctionnalités de décomposition de requêtes en des sous-requêtes mono-sources; 2- de la transmission efficace des sous-requêtes locales aux sources de données; 3- de la récupération des résultats en XML depuis une interface SAX; 4- du traitement et de l'intégration des flots; et enfin 5- de l'assemblage des résultats. Les requêtes comme les sous-requêtes sont exprimées en XQuery. En outre, des descriptions de capacités de traitement sont associées aux adaptateurs de sorte que le médiateur ne puisse envoyer que des requêtes supportées aux adaptateurs. En résumé, le médiateur utilise XML pour représenter des données disparates dans un format commun et pour créer une vue unifiée de ces données. En utilisant la technologie de traitement de requêtes distribuées, le médiateur fournit aux applications les services requis pour intégrer des informations hétérogènes sur demande via des requêtes.

Cet article décrit une version du médiateur appelé XMedia. Cette version diffère de la version industrielle par certains côtés. Elle est basée notamment sur une nouvelle algèbre pour le traitement de XML appelée XAlgebra. Les contributions de cet article sont triples. Tout d'abord nous décrivons l'architecture système modulaire du médiateur XMedia. En second lieu, nous décrivons l'algorithme de traitement de requêtes, qui est basé sur des transformations de requêtes et une algèbre fonctionnant sur des tuples d'arbres XML. Un résultat important est que le médiateur est capable de traiter la plupart des requêtes sous forme de flux d'évènements XML. Troisièmement, nous utilisons un banc d'essai sur notre architecture montrant les faiblesses et les forces des composants principaux du système. Cela nous mène à de nouvelles idées pour l'optimisation de requêtes. Certaines d'entre elles devraient être intégrées dans une future version de XMedia.

Le reste de cet article est organisé comme suit. La section suivante se concentre sur les objectifs et l'architecture du *médiateur*. La section 3 décrit l'algèbre proposée (XAlgebra), une extension simple de l'algèbre relationnelle pour traiter des forêts d'arbres XML. Puis, nous donnons une vue d'ensemble de la technologie de transformation de requêtes incluse dans le médiateur XMedia à l'aide d'un exemple de requête. La section 5 présente des mesures de performance basées sur le banc d'essai TPC/R adapté à XML. Dans la section 6, nous discutons des extensions possibles du moteur d'exécution de requêtes. Nous concluons en récapitulant les contributions et en discutant de développements futurs possibles.

## 2. Vue d'ensemble et architecture

### 2.1 Intégration et interrogation de vues XML

Le médiateur XMedia est un *middleware* d'intégration de données gérant des vues XML de sources de données hétérogènes. Il suit l'approche *global-as-view* (GAV). Des vues globales sont définies par les administrateurs à l'aide de requêtes intégrant les collections locales de documents XML. Elles sont interrogées par les utilisateurs par une API Java étendant JDBC à XQuery, appelée XML/DBC. Les sources de données peuvent être de divers types : des bases de données relationnelles, des fichiers XML, des bases de données XML, des applications propriétaires, etc. Des adaptateurs spécifiques donnant les métadonnées par introspection et fournissant au moins un sous-ensemble de XQuery sur les collections exportées les encapsulent. Idéalement, un adaptateur peut supporter la définition de vues XML interrogeables en XQuery permettant de réaliser les transformations des données de la source en données XML conformes au modèle exporté.

Le médiateur respecte et s'appuie sur les normes XML, y compris XML Schéma, XQuery, les interfaces DOM et SAX. Les "XML Schemas" sont intensivement employés pour la représentation de métadonnées. En particulier, ils décrivent les sources et les vues de données situées sur n'importe quelle couche. Le typage d'une requête XQuery est vérifié à l'aide des schémas. Nous supportons actuellement la plupart des cas d'utilisation de XQuery. Nous traitons XML de façon interne sous forme de flux d'événement SAX pour des raisons d'efficacité. En effet, DOM est en général trop coûteux pour instancier des documents XML pendant le traitement. Cependant, l'utilisateur peut s'il y a lieu obtenir des arbres DOM comme résultat et nous employons parfois DOM à l'intérieur du médiateur pour conserver des documents XML pour des traitements futurs.

Les requêtes sont décomposées en sous-requêtes optimale mono-sources ainsi qu'en plans globaux de requêtes exprimés en algèbre spécifique (XAlgebra), étendant l'algèbre relationnelle pour traiter des arbres. Les requêtes sont optimisées d'une manière simple mais efficace. Des heuristiques simples sont supportées dans la version courante, mais une optimisation basée sur une estimation de coût des plans pourrait être introduite dans une version future. L'heuristique inclut l'adaptation à XML de la remontée classique en relationnel des sélections et des transformations de semi-jointure en sélections. Plusieurs algorithmes sont mis en application pour exécuter les opérateurs de la XAlgebra, le choix étant pour l'instant dicté par l'utilisateur au moyen d'indications ajoutées aux requêtes (hints).

Pour identifier les sources appropriées pour une requête et la décomposer, des métadonnées décrivant les sources sont gérées. Lorsqu'un nouvel adaptateur est enregistré sur un médiateur, les métadonnées décrivant la source sont envoyées au médiateur sous la forme d'un fichier de configuration. Ce fichier contient un document XML spécifiant le schéma pour chaque collection exposée par l'adaptateur de source. Si le schéma d'une collection n'est pas connu, un schéma par défaut est produit. Ce schéma décrit l'ensemble des chemins de la collection, c'est une forme de guide de données (*dataguide*). Les schémas de métadonnées sont maintenus dans la mémoire du médiateur et classés par source, espace de noms, collection et chemin pour un accès rapide lors du traitement des requêtes.

## 2.2 Une architecture récursive traitant des flux de données

L'architecture de médiateur est représentée figure 1. L'API XML/DBC est la seule interface avec les composants externes. On note ainsi que le médiateur envoie les requêtes aux adaptateurs via XML/DBC et obtient les résultats par XML/DBC.

Ceci permet à un médiateur de voir un autre médiateur comme un adaptateur. Les résultats sont fournis via XML/DBC par l'intermédiaire de lecteurs SAX. Ainsi, des flux d'événements sont transférés entre les médiateurs et les adaptateurs, évitant les surcoûts produits par l'allocation de structures mémoire intermédiaires. L'architecture récursive traitant les flux de données est intéressante pour les applications qui peuvent effectuer l'intégration de données à de multiples étapes de leur exécution sans trop de dégradation de performance.

Les sous-composants principaux sont : l'analyseur XQuery, le gestionnaire de métadonnées, l'évaluateur de requête, le décomposeur de requête, et le reconstructeur de résultat. Tous ces composants sont brièvement décrits ci-dessous.

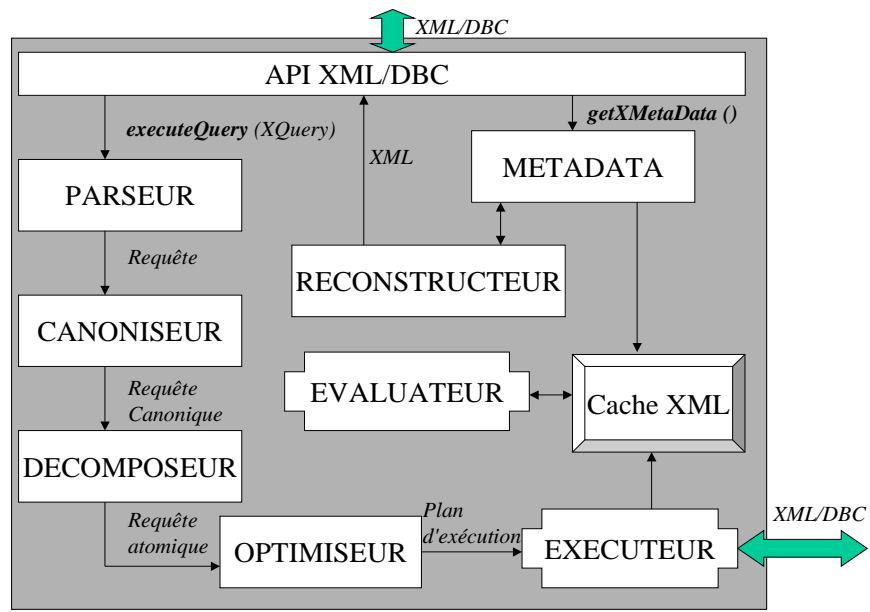

**Figure 1: Vue d'ensemble de l'architecture du médiateur**

*Analyseur de requêtes - Parseur (Parser)*

Le *parseur* analyse la requête et produit la structure de requête interne si la requête est syntaxiquement correcte et bien typée. Autrement, il renvoie une erreur documentée.

*Canoniseur (Canonizer)*

Le *canoniseur* normalise d'abord la requête et produit des requêtes sous forme canoniques. La normalisation applique les règles de transformation décrites dans (Manolescu *et al.*, 2001). Par exemple, les clauses LET sont traitées en tant que définitions temporaires de variables et éliminées. Les expressions de la forme FLWR (FLWR) sont désimbriquées si possible. Dans un second temps, le canoniseur transforme les requêtes normalisées en des requêtes simples plus un opérateur de reconstruction. Une requête simple est une requête dans laquelle toutes les expressions de retour sont des expressions simples de chemin organisées en séquence. La canonisation permet donc de décomposer une requête normalisée en requêtes simples traitant les données issues des sources et en une requête de reconstruction en mémoire. L'opérateur de reconstruction est une séquence d'éléments de construction dont les balises et les données sont des constantes ou qui proviennent d'expressions simples de chemin.

*Décomposeur (Decomposer)*

Le *décomposeur* décompose chaque requête simple en des requêtes atomiques, c.-à-d., des requêtes impliquant une seule collection globale. Il produit également un arbre de jointure (qui peut être vide) pour maintenir la dépendance entre les requêtes atomiques. Des opérateurs d'imbrication et de désimbrication peuvent également être produits pour restructurer des résultats intermédiaires. Le décomposeur identifie les métadonnées des sources de données appropriées et localise les collections. Basé sur ces informations, il traduit les requêtes atomiques sur une collection globale en une union de requêtes sur des collections locales. En particulier, il traduit les chemins globaux avec des expressions régulières dans des chemins locaux remplaçant les jokers par les chemins possibles extraits des métadonnées. En d'autres termes, il crée un premier plan d'exécution pour la requête.

*Optimiseur (Optimizer)*

Le plan d'exécution se compose d'opérateurs de la XAlgebra. Le rôle de l'*optimiseur* est de le transformer pour obtenir le meilleur plan possible. Des optimisations simples du plan de requête sont mises en oeuvre dans la version actuelle du médiateur, mais des optimisations plus complexes basées sur un modèle de coût sont prévues. Par exemple, l'optimiseur groupe les opérateurs qui référencent la même source en une seule requête pour l'expédier en une fois. Il trie également les opérateurs globaux selon l'heuristique choisie et retient la meilleure méthode de traitement (parallèle, en séquence ou en *pipeline*) pour les opérateurs globaux. Il choisit également le meilleur algorithme disponible pour chaque opérateur de l'algèbre.

*Exécuteur (Executor)*

L'*exécuteur* est responsable de l'expédition des sous-requêtes aux adaptateurs en utilisant XML/DBC. En retour, il rassemble les résultats en mémoire cache. En général, les résultats ne sont pas instanciés entièrement dans la mémoire cache mais des événements SAX sont produits et sont directement traités par l'évaluateur si possible, notamment dans les cas simples (union de résultats). Nous représentons chaque collection d'arbre XML envoyé par un adaptateur comme un XTuple, c.-à-d., un tuple contenant des références à la forêt d'arbres XML instanciée dans le cache.

*Evaluateur (Evaluator)*

En se basant sur le plan de requête, l'*évaluateur* évalue la requête globale restante et applique les opérateurs algébriques dans la mémoire principale. Les opérateurs de la XAlgebra peuvent effectuer les opérations basées sur XPath mais aussi les projections, restrictions, produits, jointures, imbrications, désimbrications, unions, intersections et différences de collections ordonnées de XTuples. Pour chaque opérateur, nous mettons en application un ou plusieurs algorithmes spécifiques. Par exemple, plusieurs algorithmes de jointures globales sont possibles. L'évaluateur peut travailler avec les collections intermédiaires entièrement stockées dans la mémoire principale, mais peut également travailler avec un flux d'évènements SAX, et de ce fait, implémenter l'évaluation en *pipeline* avec la jointure par hachage. Sont possibles également les algorithmes de jointures dépendantes demandant un XTuple à une source et interrogeant l'autre en se basant sur les résultats précédents.

*Reconstructeur (Reconstructor)*

Le *reconstructeur* applique l'opérateur de reconstruction aux résultats intermédiaires représentés comme des XTuples et produit la réponse à la requête. En d'autres termes, il imbrique et étiquette les données afin de construire le résultat final. Enfin il construit le flux d'événement SAX pour fournir les résultats à l'utilisateur.

*Gestionnaire de métadonnées (metadata manager)*

Ce composant contrôle les schémas de toutes les sources enregistrées. De plus, pour chaque source, il maintient les noms de collection avec l'ensemble associé des chemins interrogeables. L'ensemble des chemins constitue une sorte de *dataguide* donnant une vue d'ensemble de tous les chemins instanciés dans la source. Si un chemin est absent, la source ne sera pas interrogée. L'ensemble des chemins doit être fourni par l'adaptateur au médiateur lors de l'enregistrement de la source (sur commande getMetaData).

## 3. Algèbre physique

Comme décrit ci-dessus, les requêtes XQuery sont traduites dans une algèbre physique assez simple pour être susceptible d'être optimisée et exécutée. Plusieurs algèbres pour XML ont récemment été proposées (Jagadish *et al.*, 2001), (Fernandez Fernandez *et al.*, 2000), (Christophides *et al.*, 2000), (Galanis *et al.*, 2001). Notre but est d'être aussi près que possible de l'algèbre relationnelle étendue (Zaniolo, 1985), tout en permettant la manipulation des arbres et des collections ordonnées d'arbres efficacement. Nous présentons maintenant notre modèle de données relationnel étendu et son algèbre associée pour traiter des collections XML.

### 3.1 Modèle de données

Classiquement, une relation est un sous-ensemble du produit cartésien d'une liste de domaines. Avec des relations simples, les domaines sont de simples ensemble de valeurs; avec des relations d'objets, les domaines peuvent être des ensembles d'objets ou de valeurs. Nous introduisons la *XRelation*, qui peut être considérée comme un cas spécial des relations d'objets, un domaine pouvant des arbres XML. Classiquement, un arbre XML est un ensemble d'arbres ordonnés étiquetés. En outre, les liens croisés peuvent être supportés en tant qu'arêtes spéciales.

Avec les XRelations, les domaines sont des arbres XML construit sur un ensemble donné de chemins (le guide de données). Les attributs sont des XPath mettant en référence des noeuds dans les arbres XML (voir le schéma 2). Chaque attribut peut être multi-valué, c.-à-d., référencer plusieurs sous-arbres. Les XRelations sont des collections ordonnées de XTuples. Ainsi, chaque XTuple se compose d'attributs appelés XPath, dont les valeurs mettent en référence des sous-arbres dans la collection d'arbres. En conséquence, le schéma d'une XRelation est du type R (XPath+, [Path+]), où XPath définit les attributs et Path un ensemble de chemins définissant un arbre XML.

La figure 2 montre un exemple d'une XRelation composée de quatre XTuples. Le schéma de la XRelation est Example (personne/prenom, personne/nom; personne/adresse/rue, personne/titre, livre/auteur/nom, livre/date [personne/prenom, personne/nom, personne/adresse, personne/adresse/rue, personne/adresse/ville, livre/titre, livre/auteur, livre/auteur/nom, livre/date]). Un XTuple se rapporte à des noeuds et peut être considéré comme un index d'arbres XML. Le traitement par des références calculées une seule fois est beaucoup plus efficace que le traitement des arbres par navigation directe.

**Figure 2: Exemple de *XRelation***

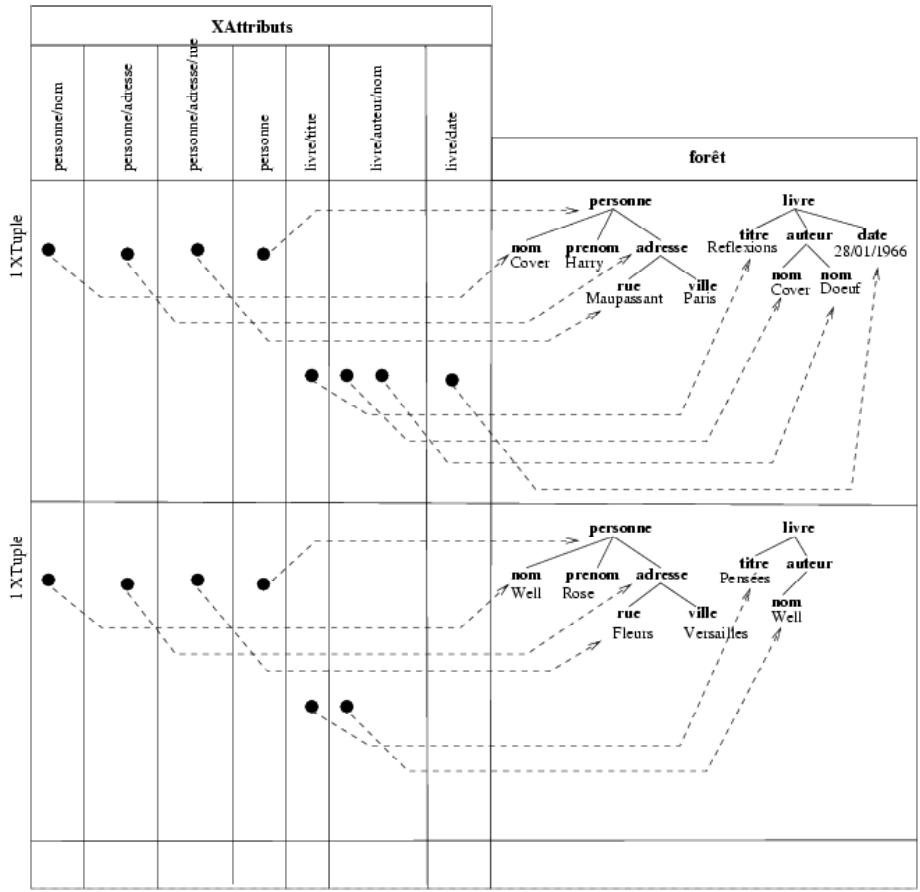

## 3.2 Opérateur de la XAlgebra

L'algèbre retenue inclut les opérations relationnelles permettant de traiter à la fois les tables, les références et la navigation dans les arbres XML. L'algèbre est une algèbre physique dans le sens où les expressions algébriques sont employées pour traiter des flux XML et que les algorithmes mettent directement en œuvre les opérateurs.

Les documents XML sont envoyés au médiateur sous forme de flux d'événements (basés sur SAX). Les *XTuples* sont créés "au vol" quand les documents XML de schémas connus sont reçus des adaptateurs. Les opérateurs non bloquants travaillent en pipeline sur le flux d'événement. Les opérateurs bloquants exigent une instanciation complète d'un flux d'entrée dans le cache. Les opérateurs n-aires non bloquants peuvent en général travailler en parallèle sur les flux d'entrée.

Tous les opérateurs de la XAlgebra reçoivent une collection de *XTuples* en entrée et renvoient une collection de *XTuples* en sortie. En général, nous modifions directement la *XRelation* en mémoire. Les opérateurs ont également des paramètres spécifiques, nous en détaillerons certains dans la suite.

Le processus d'évaluation de chaque opérateur se compose de deux étapes: une étape de préparation et une étape d'exécution. L'étape de préparation analyse le(s) *XRelation*(s) d'entrée et le(s) paramètre(s) associé(s) à l'opérateur pour déterminer quelle sera l'opération exacte à effectuer quand les *XTuples* arriveront. Par exemple, pour une opération qui exige de fusionner des arbres, l'étape de préparation détermine à quel nœud de référence le nouveau sous-arbre devra être lié et quels chemins seront mis en commun. Ainsi, l'étape d'exécution est efficace, puisque la majeure partie du traitement a déjà été faite.

*XSource*

*XSource* est l'opération de départ pour traiter une source de données XML. *XSource* prend une *XRelation* particulière de schéma (racine, [P1...Pn]) représentant une source de données comme entrée et produit une *XRelation* du schéma donné (a, b, c... [P1... Pn]), où a, b, c... sont des XPaths sur P1..., PN. Dans la pratique, *XSource* envoie une requête à une source de données et renvoie le résultat comme une *XRelation*. Pour cela, il analyse le flux SAX résultat "au vol" et produit la collection de *XTuples* en construisant les arbres et en identifiant les nœuds qui doivent être mis en référence dans la partie références des *XTuples*. *XSource* préserve l'ordre des documents des sources. C'est un opérateur non bloquant, qui peut commencer à construire les *XTuples* dès que le lecteur SAX commence à envoyer des événements.

*XProject*

*XProject* généralise la projection classique aux *XRelations*. Il prend une *XRelation* en entrée et renvoie une *XRelation* avec seulement les *XAttributs* choisis dans la table; les sous-arbres non-référencés sont également enlevés. Dans la pratique, il traite la partie référence du *XTuple* pour déterminer si le *XAttribut* doit être gardé. S'il ne doit pas être gardé, il supprime la référence et supprime les chemins non référencés dans la partie arbre. La figure 3 illustre un exemple d'exécution de l'opérateur *XProject*.

Tout comme *XSource*, *XProject* préserve l'ordre et est non bloquant.

**Figure 3: Exemple d'une opération utilisant XProject**

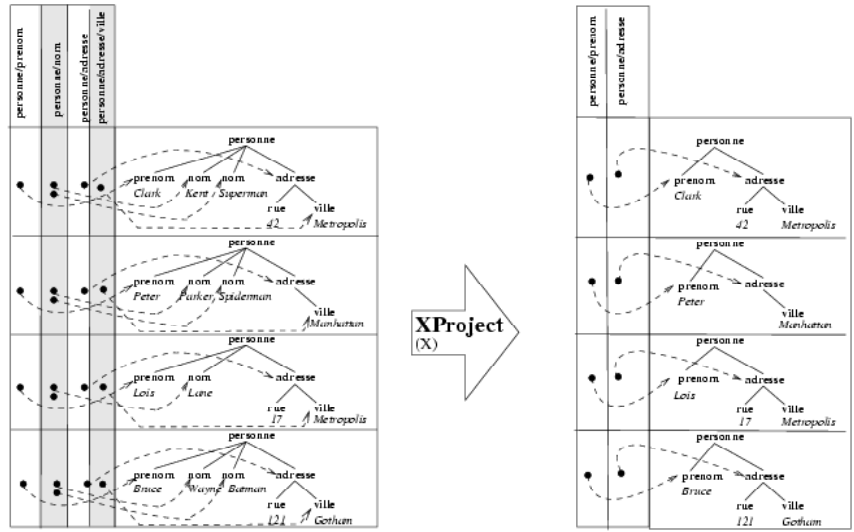

### XRestrict

L'opérateur *XRestrict* filtre chaque XTuple d'une *XRelation* sur une expression logique d'attribut, chaque attribut élémentaire comparant un attribut aux constantes ou vérifiant des contraintes d'intervalle sur un attribut. Si la condition est vraie, le XTuple est gardé. Autrement, il est enlevé avec les sous-arbres associés XML. *XRestrict* préserve l'ordre et est non bloquant.

### XProduct

L'opération *XProduct* prend deux collections comme entrée et calcule leur produit cartésien. De plus, les arbres de chaque XTuple sont fusionnés si leurs ensembles de chemin se recouvrent depuis leur racine. La figure 4 illustre une opération *XProduct*.

En général, le produit cartésien peut être calculé en pipeline mais l'ordre n'est alors pas préservé. Il est possible de préserver l'ordre d'une relation d'entrée en utilisant un algorithme de boucle imbriquée, mais l'opérateur est alors rendu bloquant. En général, ces paramètres dépendent de l'algorithme d'exécution de la même façon que l'algèbre relationnelle.

**Figure 4: Exemple d'une opération utilisant XProduct**

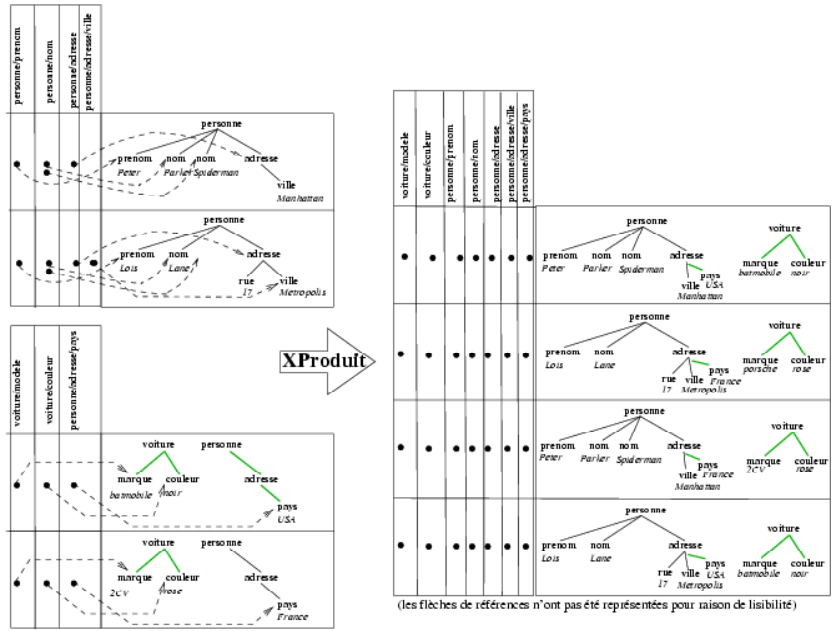

*XJoin*

    *XJoin* est la généralisation d'une jointure relationnelle. C'est un *XProduct* combiné avec un *XRestrict*. Le *XJoin* est un opérateur de base de l'algèbre physique. Plusieurs algorithmes ont été mis en application pour *XJoin* dont les boucles imbriquées, le tri-fusion et "l'interrogation d'une source avec l'autre" (jointure dépendante). Tandis que la boucle imbriquée peut être évaluée en *pipeline*, d'autres ne le peuvent pas. Seules les boucles imbriquées non évaluées en *pipeline* préservent l'ordre, mais le tri-fusion peut produire un ordre intéressant.

**Figure 5: Exemple d'une opération de jointure naturelle utilisant XJoin (sur / personne/adresse/ville)**

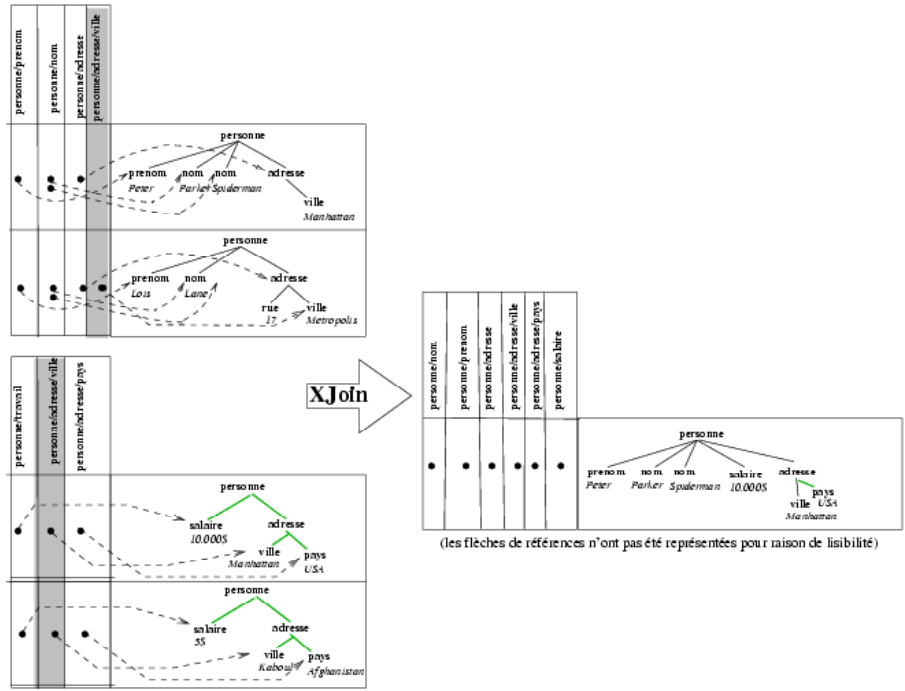

*XSort, XNest, XUnnest*

*XSort* est un opérateur simple triant une *XRelation* sur une liste donnée de XPath, par ordre décroissant ou croissant.

*XNest* applique un opérateur de groupage à une *XRelation*. Il groupe les *XTuples* qui ont les mêmes valeurs sur un ensemble d'attributs (c.-à-d., XPath) en fusionnant leurs sous-arbres communs et en insérant les branches non communes dans un arbre composé unique. Des références multi-valuées sont en général créées. C'est un opérateur coûteux qui applique d'abord un *XSort* et puis une fusion des arbres comportant des ensembles similaires de chemin. Il est illustré figure 6.

**Figure 6: Exemple d'une opération utilisant XNest**

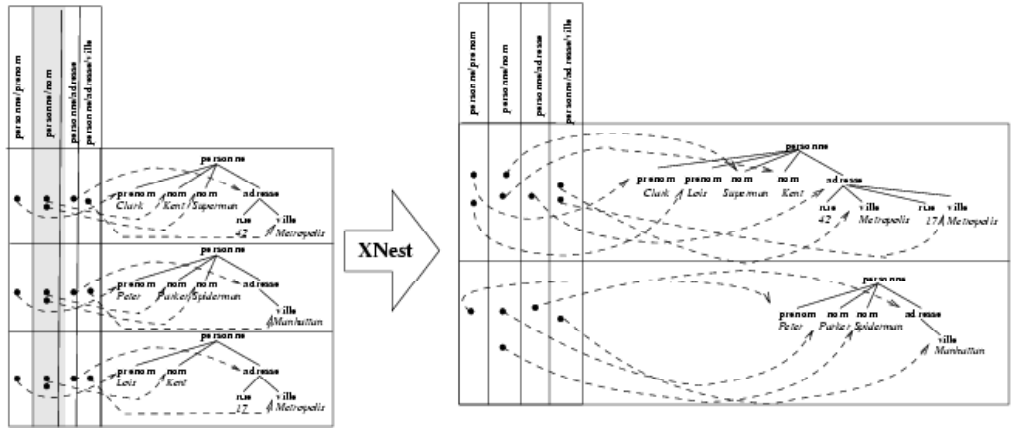

XUnnest est l'inverse de XNest:cet opérateur permet de dégrouper un attribut multi-valué d'une XRelation par rapport à un ou plusieurs attributs pivot en créant un XTuple par valeur de l'attribut multivalué en dupliquant les attributs pivot; côté arbe, il peut conduire à dupliquer des arbres pour chaque sous-arbre multivalué. Cet opérateur est un peu complexe mais est nécessaire pour assurer la complétude de l'algèbre par rapport à XQuery.

### XAggregate

Comme avec l'algèbre relationnelle étendue, le but de l'agrégation est d'appliquer une fonction MIN, MAX COUNT, AVG ou SUM à une collection de valeurs. La collection est simplement indiquée par un attribut XPath de la XRelation. Excepté avec COUNT qui compte directement le nombre de références, les fonctions s'appliquent aux valeurs référées par les attributs, qui doivent être correctement typés (numériques avec les fonctions classiques). Les XAggregate sont des opérations bloquantes ne préservant pas l'ordre.

### XReconstruct

La reconstruction est en général l'opération finale dans une expression algébrique pour éditer le flux final d'événements SAX comme résultat. Elle prend comme paramètres d'entrée une XRelation et un document XML dans lequel les valeurs sont remplacées par des attributs de la XRelation (c.-à-d., XPaths). Une instance de résultat par XTuple est alors produite. L'opération préserve l'ordre et est non bloquante. Cet opérateur est introduit lors de la canonisation des requêtes XQuery introduite ci-dessus.

### XUnion, XDifference, XIntersection

Ce sont les opérateurs ensemblistes classiques appliqués à des ensembles d'XTuples.

**4. Exemple de traitement d'une XQuery**

Comme introduit dans la section architecture, la construction d'un plan d'exécution suit les étapes suivantes :

- Normalisation et canonisation.

- Atomisation et extraction des jointures.

- Identification des sources.

- Création du plan d'exécution.

- Optimisation du plan d'exécution.

Nous allons maintenant illustrer ces étapes avec un exemple simple. Pour nos expériences, nous avons adapté le banc d'essai TPC-R à un scénario approprié pour un système fédéré semi-structuré. Nous avons pour cela groupé quelques tables ensemble pour obtenir des données arborescentes. La figure 7 décrit le schéma et la distribution des données extraites à partir du banc d'essai TPC-R.

**Figure 7 : Schéma et distribution des données**

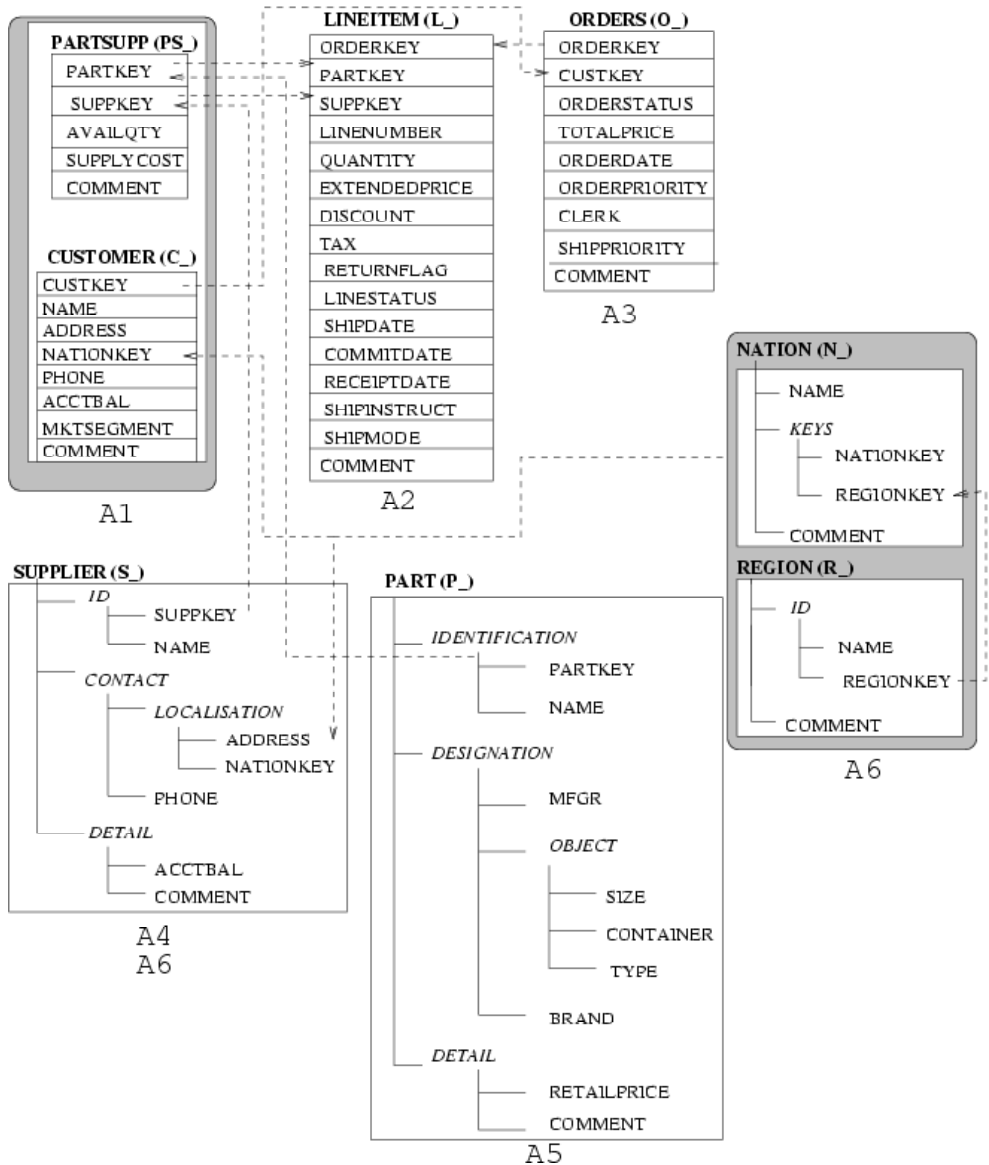

Des tables relationnelles PARTSUPP et CLIENT, LINEITEM, ORDRES sont contrôlées par les adaptateurs A1, A2 et A3 sur les collections d'un SGBD relationnel. Les collections XML arborescentes FOURNISSEUR, PARTIE, NATION et REGION sont stockées dans un SGBD XML. SUPPLIER est partitionné sur les adaptateurs A4 et A6 tandis que PARTIE est contrôlée par l'adaptateur A5. NATION et RÉGION sont contrôlés par l'adaptateur A6.

Pour illustrer les étapes de transformation de requête décrites ci-dessus, nous prenons la requête suivante:

*"Afficher pour chaque nation ayant le mot "iron" dans son commentaire, la liste de fournisseurs (nom et téléphone) qui y sont localisés avec en imbrication un partsup (partkey et supplycost) ayant une quantité disponible de plus de 45"*

La requête formelle peut être écrite en XQuery comme suit:

```
for $n in Collection("*")/nation

where contains ($n/comment, "iron")

return

<nation>

<name>$n/name</name>

<suppliers>

for $s in Collection ("*")/supplier,

    $ps in Collection ("*")/partsupp

where $s/id/suppkey = $ps/suppkey

    and $ps/availqty > 45

    and $s//nationkey = $n/nationkey

return

<supplier>$s/name</supplier>

<phone>$s/contact/phone</phone>

<partsupp>

<partkey>$ps/partkey</partkey>

<supplycost>$ps/supplycost</supplycost>

</partsupp>

</suppliers>

</nation>
```

Comme il n'y a pas de clause LET dans l'exemple, la requête est désimbriquée directement. En appliquant une règle similaire à celle définie dans (Manolescu *et al.*, 2001), la requête est désimbriquée en sous-requêtes qui sont ensuite purgées de tout balisage de reconstruction. Nous les appelons requêtes élémentaires. Puis la requête de reconstruction est générée: il s'agit simplement du document XML à retourner avec des expressions XPath au lieu des constantes.

| Requête canonisée |
|---|
| *Requête élémentaire 1* |
| **let t1 ::= for** $n **in** Collection("*")/nation |
|   **where contains** ($n/comment, "iron") |
|   **return** ($n/nationkey, $n/name) |
| *Requête élémentaire 2* |
| **let t2 := for** $t **in** $t1 |
| **for** $s **in** Collection ("*")/supplier |
| **for** $ps **in** Collection ("*")/partsupp |
| **where** $ps/availqty > 45 |
| **return**    ($s/contact/localisation/nationkey,    $s/id/suppkey,    $s/name, $s/contact/phone, ($ps/suppkey , $ps/partkey, $ps/supplycost,)) |
| *Requête de reconstruction* |
| \<nation\> |
|   \<name\>$t1/name\</name\> |
|   \<suppliers\> |
|   \<supplier\>$s/name\</supplier\> |
|   \<phone\>$s/contact/phone\</phone\> |
|       \<partsupp\> |
|       \<partkey\>$ps/partkey\</partkey\> |
|       \<supplycost\>$ps/supplycost\</supplycost\> |
|   \</partsupp\> |
|   \</suppliers\> |
|   \</nation\> |

La phase d'atomisation extrait de la requête élémentaire le maximum de sous-requêtes pour chaque collection logique avec les restrictions associées et les autres opérateurs unaires comme les tris ou les agrégats. Elle produit également la condition de jointure finale éventuellement suivie d'agrégats et de tris. Elle se termine généralement en générant un opérateur d'imbrication pour obtenir le XTuple résultat correctement imbriqué pour reconstruire les documents XML finaux. Dans notre cas simple avec seulement des restrictions et des jointures, nous obtenons trois requêtes atomiques et deux jointures suivi d'une imbrication. Elles peuvent être exprimés suivant la syntaxe XQuery de la façon suivante :

| Requête décomposée |
|---|
| *- Requête atomique t1* |
| **let t1 ::= for** $n **in** Collection("*")/nation |
| **where contains** ($n/comment, "iron") |
| **return** ($n/nationkey, $n/name) |
| *- Requête atomique t2* |
| **let t2 := for** $s **in** Collection ("*")/supplier |
| **return**    ($s/contact/localisation/nationkey,    $s/id/suppkey,    $s/name, $s/contact/phone) |
| *- Requête atomique t3* |
| **let t3 := for** $ps **in** Collection ("*")/partsupp |
| **where** $ps/availqty > 45 |
| **return** ($ps/suppkey , $ps/partkey, $ps/supplycost) |
| *- Requête globale* |
| **for** $n **in** t1, $s **in** t2, $ps **in** t3 |
| **where** |
| $s/id/suppkey = $ps/suppkey |
| **and** $s/ /nationkey = $n/nationkey |
| **return** |
| ($n/name, ($s/name, $s/contact/phone, ($ps/partkey, $ps/supplycost))) |

La requête est encore analysée afin d'identifier les sources de données contribuant au résultat. Les métadonnées décrivant chaque source enregistrée sont employées pour déterminer la pertinence de la source et pour remplir les jokers  de XPath. Notons qu'une source peut gérer plusieurs collections et qu'une collection peut être trouvée sur plusieurs sources. Pour les requêtes atomiques T1, T2 et T3, nous obtenons:

| Requête atomique | Chemins utilisés | sources |
|---|---|---|
| t1 | Collection("NATION")/nation/comment<br><br>Collection("NATION")/nation/nationkey<br><br>Collection("NATION")/nation/name | A6 |
| t2 | Collection("SUPPLIER")/<br>supplier/contact/localisation/nationkey<br>Collection("SUPPLIER")/supplier/id/suppkey<br>Collection("SUPPLIER")/supplier/id/name<br>Collection("SUPPLIER")/supplier/contact/phone | A4, A6 |

| t3 | Collection("PARTSUPP")/availqty | A1 |
| | Collection("PARTSUPP")/suppkey | |
| | Collection("PARTSUPP")/partkey | |
| | Collection("PARTSUPP")/supplycost | |

Le plan d'exécution peut maintenant être construit en termes d'opérateur de la XAlgebre. Pour chaque requête atomique, un opérateur XSource est créé. Son rôle est d'envoyer la requête à l'adaptateur et d'obtenir le résultat sous la forme de XTuple. La requête globale est utilisée comme moyen de recomposer l'arbre de jointure et l'opérateur d'imbrication. En conclusion, l'opérateur XReconstruct est ajouté pour produire le résultat XML correct. Le plan d'exécution proposé pour la requête d'exemple est représenté figure 8. Celui-ci devra évidement encore être optimisé.

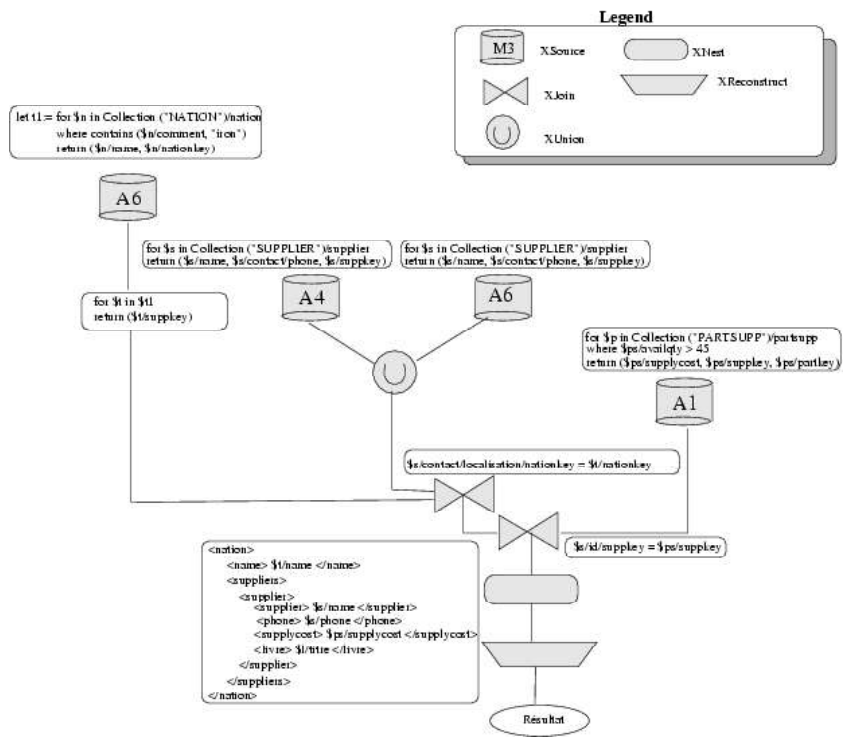

**Figure 8: Plan d'exécution proposé pour la requête**

L'arbre algébrique peut être optimisé en utilisant des règles traditionnelles de l'algèbre relationnelle imbriquée : exécution des restrictions au début, remontée des projections, ordonnancement des jointures, choix du meilleur algorithme pour chaque opérateur. Cette dernière optimisation exige des indications de l'utilisateur (*hints*) ou un modèle de coût. Nous discuterons de ce point dans la suite.

## 5. Mesures de performance

Pour comprendre où se trouvent les goulots d'étranglement du système et déterminer les optimisations qu'il serait utile d'étudier, nous avons expérimenté avec une version bêta du système industriel. Dans cette section, nous décrivons quelques résultats de nos expériences montrant les surcoûts induits par chaque composant de l'architecture.

### 5.1 Architecture d'évaluation

Nous avons utilisé une architecture client/serveur avec deux machines serveurs. Le processeur de la machine cliente est un Celeron 600 Mégahertz avec 64 Mo de RAM. Les machines serveurs sont toutes les deux des Pentium 4 à 1,6 Gigahertz avec 256 Mo de RAM. Le réseau est à 10 Mbits/seconde. Le système d'exploitation de ces trois machines est Linux. 2.4.

Pour comparer les diverses architectures, nous avons employé différents arrangements de médiateurs et d'adaptateurs, comme représenté sur la figure 9. M0, M1, M2, M3 et M4 sont des médiateurs. Ils sont tous lancés sur l'ordinateur client. A1, A2, A3 sont des adaptateurs sur les bases de données relationnelles. A4, A5, A6 sont des adaptateurs sur une base de données semi-structurée (c'est en effet le *repository* e-XML de e-XMLMedia). A7 est un adaptateur sur une base de données relationnelle qui contient exactement toutes les données d'A1, A2 et A3. M1 est relié aux médiateurs tandis que tous les autres sont reliés aux adaptateurs. Ceci est rendu possible car les médiateurs et les adaptateurs ont les mêmes interfaces.

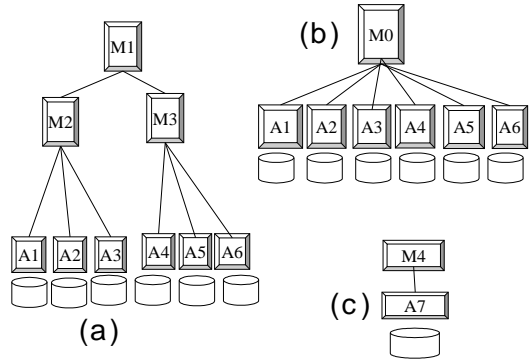

**Figure 9: Comparaison d'architectures de médiation**

### 5.2 Surcoût de médiation

Pour évaluer les surcoûts du médiateur, nous considérons un ensemble de requêtes du banc d'essai modifié de TPC/R présenté ci-dessus. La requête simple suivante:

**for** $O in **collection**("ORDERS")

**where** $O/orderkey $< N$

**return** <result> <O>$O/comment</O></result>

est exécutée successivement sur le médiateur M0, le médiateur M4 et l'adaptateur A3. $N$ varie de 1 à 3000 pour obtenir des résultats de différentes tailles. De cette façon, nous pouvons comparer le surcoût d'un médiateur sur un médiateur, et d'un médiateur sur un adaptateur. Pour comparer avec un accès direct à l'adaptateur A3, tous les objets ORDERS sont gérés par l'adaptateur A3. La figure 10 montre le temps d'exécution en fonction du nombre de documents résultats pour chaque type d'exécution.

**Figure 10: Coût d'une exécution sur M0, M1 et A3**

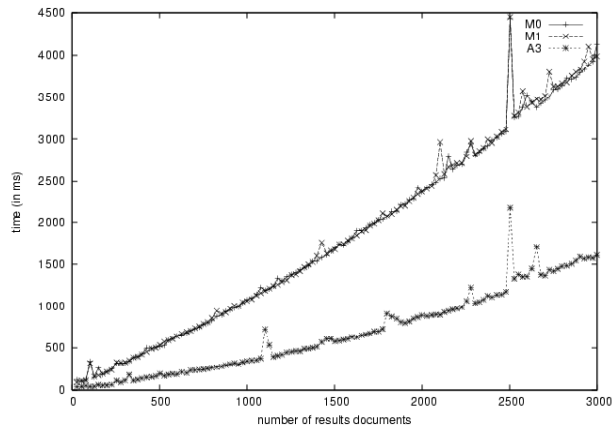

La figure montre que la période d'exécution de la requête sur M1 accédant à M2 puis A3 diffère de moins de 10% de la période d'exécution de M0 accédant directement à A3. Ceci démontre la valeur de notre architecture récursive et en général les faibles surcoût introduits par le médiateur pour des requêtes simples. Notons cependant qu'exécuter la requête directement sur l'adaptateur prend environ la moitié du temps. Ceci est dû au temps requis pour transférer et convertir les données en XML.

### 5.3 Coût par étapes

Comme détaillé ci-dessus, le traitement d'une requête suit les étapes ci-dessous:

Analyse de la requête qui transforme la requête XQuery sous format interne.

Construction d'arbre algébrique qui normalise, canonise, et atomise la requête et construit finalement l'arbre algébrique.

Initialisation de l'exécution établissant la connexion aux adaptateurs et obtenant le premier XTuple.

Exécution locale de la requête sur l'adaptateur comprenant l'envoi de la requête à l'adaptateur, l'obtention du résultat par XML/DBC dans le format SAX et la transformation du flux SAX en XTuple.

Exécution globale de la requête et reconstruction, c.-à-d., le traitement des XTuples par l'arbre algébrique pour renvoyer le résultat.

Les étapes 1, 2 et 3 composent la phase d'initialisation de traitement de requête.

Les temps passés pour la phase d'initialisation et pour les étapes 4, 5, et pour le traitement complet sont décrits dans la figure 11. L'étape d'initialisation est négligeable devant les autres temps. Le temps total est encore approximativement le double du temps pris par l'adaptateur. L'évaluation sur l'adaptateur se compose de :

- La transformation de la requête en SQL.

- L'exécution de la requête sur la base de données (Oracle).

- La récupération des tuples et la transformation en documents XML.

Comme les résultats sont mesurés sur une base de données à chaud, les tuples sont dans le cache et les requêtes SQL sont exécutées dans la mémoire centrale. Ceci confirme que le temps dominant est la construction et la communication de documents XML.

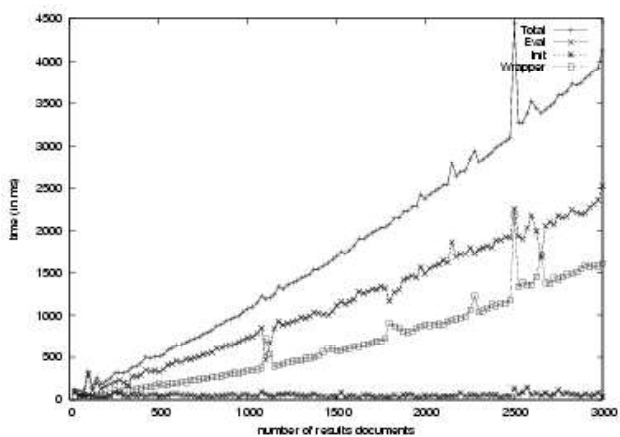

**Figure 11: Temps d'exécution pour chaque phase**

Sur la figure 12, nous détaillons le temps d'initialisation entre le moment d'analyse de la requête, le moment de production du plan d'exécution et le moment d'obtention du premier résultat. Tous ces temps sont courts. Le temps d'analyse de la requête est très court (<10 ms.). La production du plan d'exécution prend un peu plus de temps (<15ms.). L'obtention du premier résultat a besoin d'un peu plus de temps, prouvant encore que le temps d'échange est dominant.

**Figure 12: Temps d'exécution pour les différentes étapes de la phase d'initialisation**

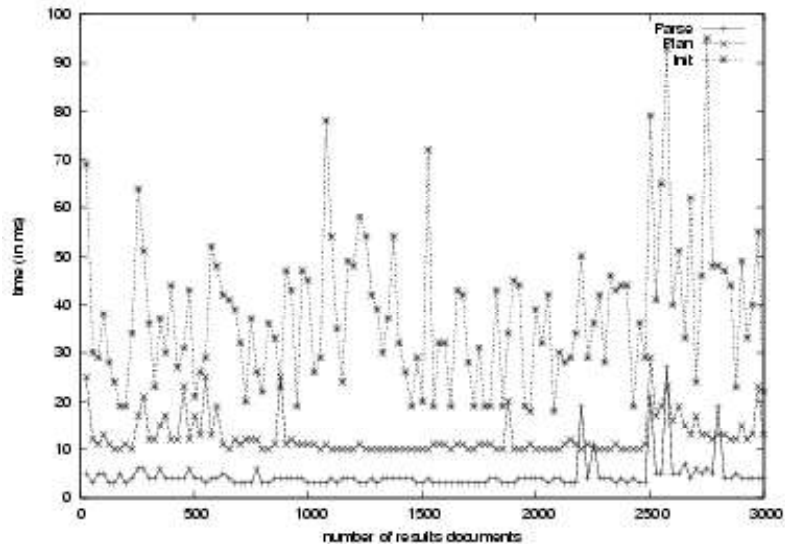

### 5.4 Jointure intersites

Nous soumettons maintenant un ensemble de requêtes qui exécutent une jointure entre deux tables. Soit la requête suivante :

**for** $L **in collection**("LINEITEM")

**for** $O **in collection**("ORDERS")

**where** $O/orderkey = $L/orderkey

 **and** $L/orderkey < *N*

**return** <result>

<lcom>$L/comment</lcom>

<ocom>$O/comment</ocom>

 </result>

Comme précédemment, *N* varie de 1 à 3000 pour faire varier la sélectivité. Nous évaluons d'abord la requête sur le médiateur M4, puis sur le médiateur M2. Dans le premier cas, la jointure est exécutée par la source de données (Oracle) dans la mémoire du serveur; dans le deuxième cas, la jointure est exécutée sur le médiateur et des tuples XML sont transférés sur le réseau. Encore une fois, le résultat (voir figure 13) prouve que le temps de transfert est dominant. Il prouve également que la jointure intersite est une opération coûteuse qui devrait être poussée vers l'adaptateur si possible.

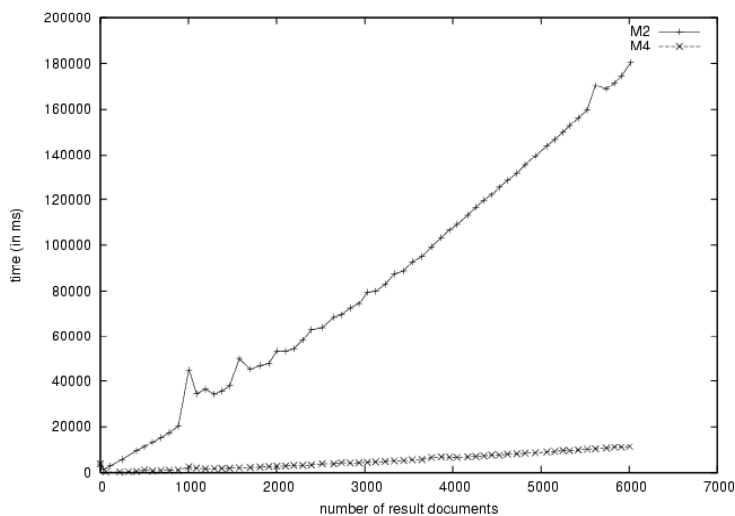

**Figure 13: Temps d'exécution sur M2 et sur M4**

## 6. Améliorations possibles

Les résultats des expériences rapportées ci-dessus, démontrent le coût élevé de communication pour l'échange de documents XML entre les adaptateurs et les médiateurs. C'est le premier point à améliorer. Nous proposons plusieurs améliorations qui devraient réduire ce coût.

### 6.1 Compression XML et transfert brut

Transférer des documents XML entre les adaptateurs et les médiateurs semble être coûteux. Chaque XTuple est codé dans un message XML et envoyé sur le réseau. Le message XML est alors analysé au niveau client et transformé en interne en un descripteur de XTuple et en arbres XML au fil du flux d'évènements. Ainsi, le nombre de messages est important et la durée de la transformation est longue. On peut arguer du fait que notre réseau est lent (10 Mbits), mais ce n'est pas suffisant pour expliquer les résultats.

Pour gagner en nombre de messages, nous pourrions employer le transfert en masse, et envoyer plusieurs messages dans un bloc. Le nombre de messages par bloc devrait être accordé de telle sorte que le *pipeline* sur le client puisse continuer à travailler sans à-coup. Néanmoins, ceci n'empêche pas l'analyse et la transformation des messages très longs. C'est de toute façon inhérent à XML et ceci dégradera toujours les exécutions.

Une solution est d'employer un format compressé pour transférer les XTuples. Les schémas de XTuples sont connus par le client et le serveur sous la forme d'une liste de chemins. Les types de valeurs (feuilles des arbres XML) sont également connus par des schémas XML. Ainsi, un mécanisme évident de compression consiste à envoyer un XTuple comme une séquence d'identifiants de chemin (16 bits sont suffisants) suivie de la valeur de feuille codée selon son type. L'analyse sera alors une tâche évidente. Cependant, nous nous éloignons alors de la philosophie de XML et de la généralité du mécanisme de communication. Bien que ce soit un peu contraire aux principes de XML, nous croyons qu'un dispositif de compression permettant d'économiser du temps d'analyse est crucial.

### 6.2 Algorithmes implantant les opérateurs

La version testée du médiateur utilise un algorithme simple de jointure (boucles imbriquées optimisées). Il est évident que d'autres algorithmes devraient être considérés, pour la jointure notamment, mais aussi pour d'autres opérateurs (par exemple, pour l'imbrication qui est assez complexe). Implémenter la jointure dépendante, c.-à-d., une jointure lisant une *XRelation* et en interrogeant l'autre avec la valeur lue, pourrait être utile pour gagner en nombre des messages en cas de résultat de faible cardinalité. La jointure par tri-fusion et la jointure par hachage pourraient également être utiles. Ainsi, nous intégrons actuellement une bibliothèque d'algorithmes pour chaque opérateur de la XAlgebra. Le problème est alors de choisir le meilleur plan. Une solution sophistiquée consiste à développer un modèle de coût.

### 6.3 Modèle de coût

La solution classique pour choisir le meilleur plan d'exécution est de comparer les coûts des différents plans en utilisant un modèle de coût. Nous proposons un modèle de coût fortement inspiré de DISCO (Tomasic *et al.*, 1996). Le médiateur est muni d'un modèle de coût générique dérivé d'un modèle de coût relationnel étendu avec la manipulation d'arbre. Chaque adaptateur peut alors exporter des statistiques et des formules détaillées de coût vers le médiateur. Le modèle générique de coût est généralement employé avec des exportations de statistiques (pour évaluer des cardinalités), et les formules spécifiques exportées par un adaptateur peuvent surcharger les formules génériques. Cette approche donne un cadre pour calculer le coût global d'un plan de requête intégrant l'information locale des sources.

Pour communiquer leur modèle de coût au médiateur, un adaptateur emploie un langage de modèle de coût. Dans un environnement XML, le langage de coût doit être défini en XML. Comme les formules et les définitions de statistiques emploient beaucoup de notations mathématiques, nous avons construit une proposition de langage de coût sur MathML. MathML est une spécification du W3C pour coder en

XML la représentation ou la structure d'un objet mathématique. Seules les informations structurelles sur un objet mathématique sont intéressantes dans notre cas. Les avantages d'employer le format MathML pour décrire des formules de coût sont triples: il est entièrement en XML, il supporte des formules générales, et des logiciels courant de calcul peuvent être employés pour calculer les formules.

Les paramètres utilisés pour l'évaluation d'un modèle de coût sont des statistiques relatives au système (statistiques système) et des statistiques relatives aux données (statistiques de données). Pour des données semi-structurées, quelques autres paramètres système devraient être définis, comme la comparaison entre deux valeurs typées, la comparaison entre deux arbres, la navigation dans un arbre (suivi de pointeurs). Les statistiques de données dépendent des données et des collections des données contenues dans la source. Les statistiques classiques de données utilisées sont: la cardinalité d'une collection, la distribution d'un attribut dans une collection, les valeurs minimum et maximum prises par un attribut. Pour des données semi-structurées, on doit ajouter certains paramètres tels que la profondeur et la largeur moyennes des arbres dans une collection. Une telle information pourrait être dérivée des schémas XML.

Un modèle de coût de médiation dépend des paramètres système et des paramètres de données choisis. Une ou plusieurs formules sont définies afin de calculer le coût d'évaluation d'une requête dans ce système (grosse granularité) ou un attribut dans un opérateur particulier (granularité fine). Les formules pour les granularités les plus fines sont spécifiques aux sources et peuvent être exprimées avec des paramètres spécifiques. Les formules pour les granularités les plus grandes se composent de la cardinalité, du coût total et du coût d'exécution.

En résumé, développer un modèle générique complet de coût avec surcharge des adaptateurs est possible dans un médiateur XML. Des formules de coût peuvent être échangées en XML. Un modèle de coût est nécessaire pour choisir les meilleurs plans d'exécution, basés sur des estimateurs des coûts de communication et des coûts de traitement.

### 6.4 Capacité des adaptateurs

Dans la version décrite du médiateur, les capacités de source sont prises en considération par catégorie. Nous supportons trois catégories des sources: Des sources XQuery, des sources SQL, et des fichiers XML. Fondamentalement nous poussons les requêtes XQuery aux sources XQuery, SQL de base aux sources SQL, et les sélections aux fichiers gérés par un filtre. Cela est insuffisant pour prendre en compte des capacités de traitement détaillées des sources. Pour aller plus loin et tenir compte des capacités de traitement détaillées des sources au niveau du médiateur, une description précise des capacités de traitement de l'adaptateur est exigée. Ceci peut être fait globalement pour un adaptateur en envoyant un fichier XML associé aux métadonnées détaillant quels opérateurs XML sont autorisés sur toutes les collections ou spécifiquement sur une collection. Les règles les plus spécifiques prévalent toujours.

### 6.5 Cache sémantique

Une autre manière de réduire la transmission de messages est de mettre en œuvre un cache sémantique au niveau du médiateur. Les XTuples répondant à une requête exécutée sur le médiateur peuvent être conservés dans un tel cache. Le format XML n'est pas approprié car trop volumineux; nous employons plutôt le format comprimé présenté ci-dessus. Ainsi une table des requêtes exécutée ordonnée par horodate d'exécution avec résultats associés devrait être maintenue dans le cache. Celle-ci serait employée pour répondre à de nouvelles requêtes. Naturellement, la mise à jour sur les données des sources ne peut être prise en considération sans mécanisme de remontée d'événements. Ainsi, le cache sémantique est seulement utilisable sous certaines collections de documents XML non mis à jour fréquemment. Il est cependant de grande utilité dans le cas de sources lentes, par exemple, les sources web.

Avec le cache sémantique, une nouvelle requête devrait d'abord être vérifiée par le cache pour déterminer si il peut répondre totalement ou en partie à la requête. Si oui, la requête est divisée en deux parties (une partie peut être nulle): une requête locale qui peut être exécutée par le cache et une requête de source qui doit être exécutée par les sources distantes. Les deux résultats doivent être correctement assemblés. Ceci peut être fait en comparant les formes canoniques des arbres algébriques associés à la requête à celle de chaque requête du cache. Si l'une est un sous-ensemble de l'autre, le cache peut être employée pour traiter une partie de la requête. L'arbre algébrique de requête doit être élagué pour remplacer la partie commune par un appel aux XRelations du cache. Employer un cache sémantique XML pour XQuery est un sujet complexe qui doit être encore étudié, mais qui pourrait être très bénéfique aux performances.

## 7. Conclusion

Nous avons présenté le système XMedia permettant d'interroger en XQuery des vues intégrées de données hétérogènes. Une première version du système a été développée à l'université à la fin des années 90, puis transférée à l'industrie de 2000 à 2002 où elle a été complètement remodelée. La deuxième version est commercialisée et a eu plusieurs applications utilisées ou planifiées, notamment dans le domaine du tourisme, de la santé, et de la pharmacie. Elle va être disponible en logiciel libre. Actuellement, un nouveau projet de recherche est en cours visant à développer un médiateur amélioré, qui devrait tenir compte des leçons du passé.

La version décrite dans cet article a des fonctionnalités uniques. Les requêtes XQuery sont compilées dans des plans d'exécution exprimés en algèbre relationnelle étendue capable de traiter les arbres XML en *pipeline*. Le traitement des requêtes est clairement divisé en étapes. Nous avons isolé l'étape de réécriture de requête de l'étape de décomposition qui produit des arbres algébriques traitant des sources de données localisées. La localisation des collections est effectuée en utilisant les métadonnées sous la forme de schémas XML. L'étape d'optimisation exige un modèle de coût pour être entièrement efficace. Des techniques d'indicateurs (*hints*) ont été intégrées dans la version industrielle.

Les mesures d'exécution démontrent la validité de l'approche mais le coût de transfert des fichiers XML à partir des adaptateurs aux médiateurs semble être excessif. Plusieurs améliorations possibles qui devraient être en partie mises en œuvre dans la version future ont été suggérées. Nous voudrions également séparer plus clairement la phase compilation de requêtes éventuellement paramétrées de la phase exécution. Il serait aussi possible de développer une X-machine virtuelle distribuée plus efficace pour traiter des expressions de XAlgebra sur des flux XML.

## 8. Bibliographie